\documentclass{jkas}

\def\year{} 
\def\month{} 
\def\volume{} 
\def\issue{} 
\def\beginpage{} 
\def\endpage{} 
\def\received{} 
\def\accepted{} 
\date{Received \received; accepted \accepted}

\newcommand*{\aj}{AJ}
\newcommand*{\apj}{ApJ}
\newcommand*{\nat}{NATURE}
\newcommand*{\mnras}{MNRAS}
\newcommand*{\aap}{A\&A}

\newcommand*{\apjl}{ApJL}
\newcommand*{\apjs}{ApJS}

\newcommand*{\teff}{$T_{\rm eff}$}
\newcommand*{\logg}{$\log~g$}

\newcommand*{\feh}{[Fe/H]}
\newcommand*{\cfe}{[C/Fe]}

\newcommand*{\kms}{km s$^{-1}$}

\newcommand*{\zmax}{$Z_{\rm max}$}
\newcommand*{\rmax}{$r_{\rm max}$}
\newcommand*{\rmin}{$r_{\rm min}$}

\newcommand*{\vphi}{$V_{\rm \phi}$}
\newcommand*{\vtheta}{$V_{\rm \theta}$}

\newcommand*{\rsun}{$R_\odot$}
\newcommand*{\zsun}{$Z_\odot$}

\usepackage{color}
\usepackage{hyperref}
\usepackage{lineno}
\usepackage{natbib}
\hypersetup{
    unicode=false,          
    pdftoolbar=true,        
    pdfmenubar=true,        
    pdffitwindow=true,     
    pdfstartview={FitH},    
    pdftitle={},    
    pdfsubject={Astronomy},   
    pdfcreator={dvipdf},   
    pdfproducer={dvipdf}, 
    pdfnewwindow=true,      
    colorlinks=true,       
    linkcolor=red,          
    citecolor=blue,        
    filecolor=magenta,      
    urlcolor=cyan,           
    breaklinks=true,
    linktocpage
}

\title{Diverse Chemo-Dynamical Properties of Nitrogen-Rich Stars Identified From Low-Resolution Spectra}

\author[1]{Changmin Kim}
\author[2,3]{Young Sun Lee}
\author[3]{Timothy C. Beers}
\author[2]{Young Kwang Kim}

\affil[1]{Department of Astronomy, Space Science and Geology, Chungnam National University, Daejeon 34134, Korea}
\affil[2]{Department of Astronomy and Space Science, Chungnam National University, Daejeon 34134, Korea}
\affil[3]{Department of Physics and Astronomy and JINA Center for the Evolution of the Elements, University of Notre Dame, IN 46556, USA}

\def\corrauthor{
Young Sun Lee (youngsun@cnu.ac.kr)
}

\def\runningauthor{
Kim et al.
}
\def\runningtitle{
}

\def\keywords{
Method: data analysis -- technique: spectroscopy --
Galaxy: halo -- stars: abundances
}

\def\abstracttext{
The second generation of stars in the globular clusters (GCs) of the Milky Way (MW)
exhibit unusually high N, Na, or Al, compared to typical Galactic halo stars
at similar metallicities. The halo field stars enhanced with such elements are believed
to have originated in disrupted GCs or escaped from existing GCs. We identify such
stars in the metallicity range --3.0 $<$ [Fe/H] $<$ 0.0 from a sample of $\sim$ 36,800
giant stars observed in the Sloan Digital Sky Survey and Large Sky Area
Multi-Object Fiber Spectroscopic Telescope survey, and present their dynamical properties.
The N-rich population (NRP) and N-normal population (NNP) among
our giant sample do not exhibit similarities in either in their metallicity distribution
function (MDF) or dynamical properties. We find that, even though
the MDF of the NRP looks similar to that of the MW's GCs in the range of [Fe/H] $<$ --1.0,
our analysis of the dynamical properties does not indicate
similarities between them in the same metallicity range, implying that the escaped members from
existing GCs may account for a small fraction of our N-rich stars, or
the orbits of the present GCs have been altered by the dynamical friction of the MW.
We also find a significant increase in the fraction of N-rich stars
in the halo field in the very metal-poor (VMP; [Fe/H] $<$ --2.0) regime, comprising up
to $\sim$ 20\% of the fraction of the N-rich stars below [Fe/H] = --2.5, hinting that partially
or fully destroyed VMP GCs may have in some degree contributed to the Galactic
halo. A more detailed dynamical analysis of the NRP reveals that our sample
of N-rich stars do not share a single common origin. Although a substantial fraction
of the N-rich stars seem to originate from the GCs formed in situ, more than 60\% of
them are not associated with those of typical Galactic populations,
but probably have extragalactic origins associated with
Gaia Sausage/Enceladus, Sequoia, and Sagittarius dwarf galaxies, as well
as with presently unrecognized progenitors.
}

\begin{document}

\jkashead 

\section{Introduction} \label{sec1}

The Milky Way (MW) hosts about 170 globular clusters (GCs)
(e.g., \citealt{gratton2004,baumgardt2021,vasiliev2021}, and references
therein). Most GCs are known to exhibit signatures of multiple stellar
populations (e.g., \citealt{carretta2009,carretta2010,milone2015,piotto2015}),
which are generally divided into two groups of stars according to
differences in their individual elemental abundances: first-generation (FG)
and second-generation (SG) stars. Numerous spectroscopic studies of GCs have
revealed that the FG stars in GCs overall follow the chemical-abundance
patterns of the Galactic field stars at similar
metallicities. On the contrary, the SG stars exhibit enhancements
of N, Na, and Al, while C, O, and Mg are depleted (see \citealt{bastian2018}
for a detailed review), establishing anti-correlations between C and N, O and Na,
and Mg and Al \citep{carretta2009,carretta2010,carretta2016,martell2016,tang2020,horta2021}.

A large number of chemical-abundance studies have reported
that some fraction of the Galactic halo stars exhibit anomalous abundances of
these same elements \citep{carretta2010,fernandez2016,martell2016,pereira2019},
as well as enhancements of slow neutron-capture
elements (e.g., \citealt{majewski2012,hasselquist2016,pereira2017}),
which are characteristics frequently observed among the SG
stars in GCs \citep{carretta2009,carretta2016,pancino2017,schiavon2017a,tang2021,fernandez2022}.
These studies argued for a possible connection
in terms of their origin between the SG stars and field halo stars with large
enhancements of N, Na, or Al. Such field stars are now believed to have originated from
tidally disrupted GCs \citep{carretta2009,carretta2010,carretta2016,fernandez2016,
fernandez2017,fernandez2019,fernandez2022,martell2016,schiavon2017a,horta2021,tang2021}.
A recent study of the tidally disrupting GC Palomar 5 finds N-rich stars both within
and beyond its tidal radius, strongly supporting the idea that the stars enhanced with
N, Na, or Al in the Galactic halo field have GC origins \citep{phillips2022}.

The existence of such stars in the halo has profound implications on the assembly history
of the MW, as GCs could contribute significantly to building up the
baryonic mass of the MW \citep{carretta2010,martell2016,schiavon2017a,lee2019,
hughes2020,thomas2020,wan2020,lim2021}, through their dissolution and/or
evaporation (e.g., \citealt{elmegreen2010,kruijssen2011}).
Consequently, understanding the fraction and nature of stars associated with the
disrupted population from the GCs can provide
important constraints on models for the assembly history of the MW. Among
the stars with the GC origin, the abundance patterns of the FG stars are indistinguishable from
the field halo stellar population, hence stars with the chemical abundances of the SG stars are
used as tracers of the members of dissolved GCs. That is, while the SG stars reveal that partially
or totally disrupted GCs have contributed to the halo, there are $also$ the FG stars that have contributed,
but are not readily identifiable.

Upon recognizing their importance, using the abundances of N, Na,
or Al from various spectroscopic survey data, such as
the Sloan Extension for Galactic Understanding and
Exploration (SEGUE; \citealt{yanny2009,rockosi2022}),
the Apache Point Observatory Galactic Evolution Experiment \citep[APOGEE;][]{majewski2017},
and the Large Sky Area Multi-Object Fiber Spectroscopic Telescope \citep[LAMOST;][]{cui2012},
numerous studies have attempted to discover examples of stars that were once GC members, but have
been stripped from their progenitors \citep{martell2010,martell2016,fernandez2016,
fernandez2017,fernandez2019,fernandez2022,schiavon2017a,koch2019,hanke2020,horta2021,kisku2021},
in order to estimate their contribution to the mass assembly of the MW halo.
These studies predicted fractions of 2 -- 28\% of SG stars originating in GCs,
depending on the spatial extent explored in the surveys. Their contribution appears
to become larger in the Galactic bulge than the halo \citep{schiavon2017a,horta2021}.
These results confirm that stars with GC origins are one source of stellar constituents of the
Galactic halo and bulge populations, indicating that the GCs must have played a role
in the assembly of the MW.

The GCs in the MW originated from two different paths: in situ and ex situ.
The ex situ (or accreted) channel consists of GCs that once belonged to dwarf galaxies
which were accreted by the MW (e.g., \citealt{pfeffer2019,hughes2020}).
The in situ path produces GCs that are thought to have formed in the turbulent
disk of the MW at redshifts of z $\sim$ 2 -- 3 (e.g., \citealt{kruijssen2012}). This implies
that the GC-origin stars not only came from GCs that formed in situ, but also those accreted
from dwarf satellites \citep{fernandez2020b,fernandez2021a}. A recent study used the APOGEE
data in the Sloan Digital Sky Survey \citep[SDSS;][]{york2000} Data
Release 17 (DR17; \citealt{abdurrouf2022}) to identify 149 N-rich ([N/Fe] $\geq$ +0.5) red
giants with [C/Fe] $<$ +0.15 in the metallicity range --1.8 $<$ [Fe/H] $<$ --0.7, combined
with previously identified N-rich stars, to construct a catalog of 412 unique N-enhanced
stars \citep{fernandez2022}.
These authors carried out a dynamical analysis of the compiled N-rich stars,
and found that their origins are diverse; some may
be dissolved members of GCs, which were brought into the MW during
major merger events such as the Gaia-Sausage/Enceladus (GSE; \citealt{belokurov2018,helmi2018}).
This indicates that extragalactic GCs may take part in the assembly of the MW as well.

Despite enormous progress in recent years, the identified
chemically peculiar field stars presumably formed in GCs are not
sufficiently numerous to closely examine their diverse origins. Furthermore,
most previous studies focused on the narrow range of the metallicity (--2.0 $<$ [Fe/H] $<$ --1.0).
The identification of many more such stars, covering a wider metallicity
range, is clearly desirable, and will provide a more clear picture on
their role in the assembly of the MW.
In this regard, low-resolution spectroscopic surveys such as SDSS and LAMOST, which
provide a large dataset are excellent sources for conducting searches for N-rich
stars over a wide range of metallicity and luminosity, compared to high-resolution
spectroscopic data, which are best suited for detailed
analyses of their chemical elements. This study presents a sample of N-rich
stars identified from SDSS and LAMOST, and carries out an analysis of their
dynamical properties in order to explore the diverse origins of their host systems.

The outline of this paper is as follows. In Section \ref{sec2}, we
describe how we select the N-rich stars, followed by calculations
of the space velocities and orbital parameters in Section \ref{sec3}.
We present our findings and discuss their implications, along with those of previous studies,
in Section \ref{res}. Section \ref{sum} summarizes our conclusions.

\begin{figure*}[!t]
\centering
\includegraphics[width=0.8\textwidth]{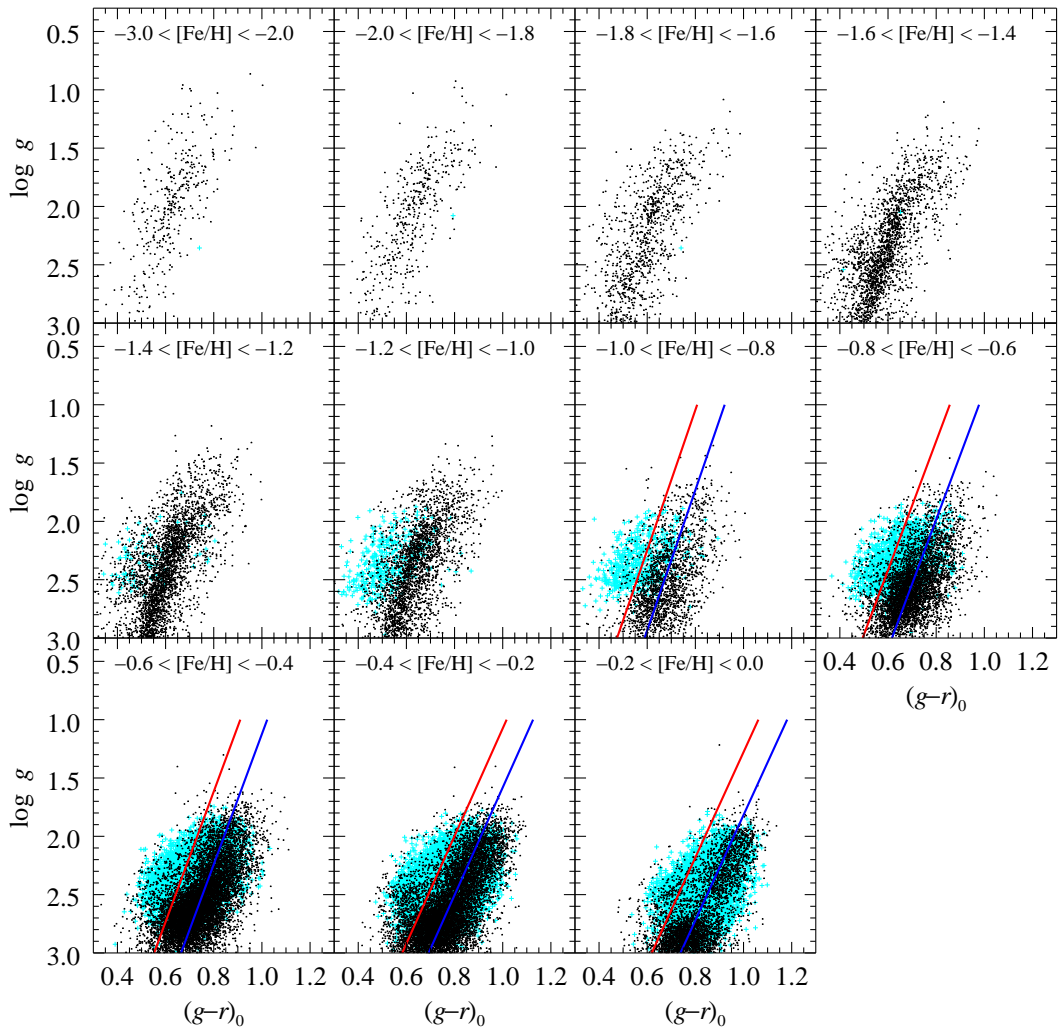}
\caption{Distribution of the selected red giant-branch (RGB)
stars in the \logg\ versus $(g-r)_{0}$ plane in different
metallicity intervals. The metallicity bin is indicated
at the top of each panel. The red lines
in the panels of the metallicity bins with [Fe/H] $>$ --1.0 are the
fiducials used to remove red-clump (RC) stars.
The cyan symbols are the RC stars identified by \citet{huang2020}.
The fiducial line is set to 2$\sigma$ away from the main locus (blue line).}
\label{fig1}
\end{figure*}

\begin{figure} 
\centering
\includegraphics[width=0.95\columnwidth]{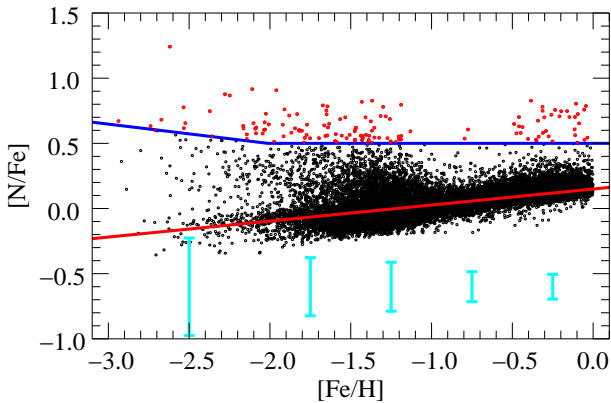}
\caption{Distribution of our selected RGB stars in the [N/Fe] versus [Fe/H] space.
The cyan error bars at the bottom indicate the typical uncertainties
of the measured [N/Fe] over bins of 0.5 dex in [Fe/H], except the left-most bin, which
is over a 1 dex bin. The blue-solid line is the dividing line to select the N-rich stars. See the text for
detailed description of deriving the line. The red dots
are the N-rich population (NRP) identified in this study, while
the black dots are defined as N-normal population (NNP).}
\label{fig2}
\end{figure}

\section{Selection of Nitrogen-rich Stars} \label{sec2}

To identify N-rich stars from the SDSS and LAMOST surveys,
we first need to determine stellar parameters (\teff, \logg, and \feh).
We applied the SEGUE Stellar Parameter
Pipeline (SSPP; \citealt{allende2008,lee2008a,lee2008b,lee2011a,
smolinski2011,lee2013}) to low-resolution spectra from SDSS and LAMOST
in order to determine the stellar parameters. Thanks to the similar spectral
wavelength range (3700 -- 9000 \AA) and resolution ($R \sim$ 1800)
of the LAMOST spectra to those of the SDSS, we were able to
derive the stellar parameters from the LAMOST spectra using
the SSPP (see \citealt{lee2015} for detailed information).
The application of the SSPP also delivered [Mg/Fe] and [C/Fe] estimates.
Then, we employed the method described by \citet{kim2022}
to derive [N/Fe] from low-resolution SDSS and LAMOST spectra.

Briefly describing the method, we normalized an observed spectrum
in the wavelength range of 3810 -- 3900 \AA, in which the CN features exist, and searched the grid of
synthetic spectra for the best-fitting model with [N/Fe] by minimizing the difference
between the normalized observed and synthetic spectra. In this process, we adopted \teff, \logg, \feh,
and \cfe\ values from the SSPP and held them fixed, and varied only the [N/Fe] value
to create a new trial synthetic spectrum by cubic spline interpolation
over the grid of the synthetic spectra to match with the observed spectrum. The measured [N/Fe] error is typically
better than 0.3 dex for signal-to-noise ratio (S/N) larger than 10 \citep{kim2022}.
The typical uncertainties of the stellar parameters (\teff, \logg, and \feh)
are about 180 K, 0.24 dex, and 0.23 dex, respectively \citep{lee2008a,lee2008b,smolinski2011},
while the uncertainty of measured [C/Fe] is $\sim$ 0.35 dex \citep{lee2013}.

To select likely GC-origin stars, we first gathered giant stars by applying
the following selection criteria: 4500 K $<$ \teff\ $<$ 5500 K, \logg\ $<$ 3.0,
--3.0 $<$ [Fe/H] $<$ 0.0, 13.0 $<$ $g_0$ $<$ 20.5,
and S/N $>$ 20. We excluded main-sequence turnoff (MSTO) stars, because they
exhibit CN and CH bands that are too weak to
reliably derive C and N abundance ratios, owing to their warm temperatures.
Thus, in this study we focus on red giant-branch (RGB) stars.
In the selection process, we first removed the stars in the plug-plates pointing to
the GCs as well as open clusters. For multiply observed stars,
we chose to include the star with highest S/N.
We also eliminated the stars with [C/Fe] $>$ +0.15, because
the SG stars of GCs typically possess low [C/Fe] \citep{martell2016,schiavon2017a}.
The removal of such stars reduces the contamination from CH stars as well \citep{kg2015}.

Another source of contamination is red-clump (RC) stars, as they
can be disguised as giants in our sample. We excluded RC
stars by application of the following two steps. The first step is to eliminate
them by cross-matching our sample of stars with the RC catalog provided by \citet{huang2020},
who compiled RC stars from LAMOST. They are indicated by cyan symbols
in Figure \ref{fig1}, which shows our RGB stars in the \logg\ versus ($g-r$)$_{0}$ plane
in different metallicity ranges, as listed at the top of each panel.
We can observe that most of them are located in the left side of the giant branch, proving
their usefulness in removing the RC stars. In the second step,
we identified the RC stars in the \logg\ versus $(g-r)_{0}$ plane.
In the panels of the metallicity region --1.0 $<$ [Fe/H] $<$ 0.0 in Figure \ref{fig1},
we drew a fiducial line (red line), which is 2$\sigma$ away from the main locus (blue line)
that follows the densest region. We considered stars in the left side of the fiducial line as RC
stars, and removed them from the list of our RGB stars. We applied this procedure only to the stars
with [Fe/H] $>$ --1.0, because most RC stars are metal-rich ([Fe/H] $>$ --1.0).
After these steps, we are left with a sample of about 36,800 RGB stars
with available [N/Fe] estimates.

To distinguish the N-enhanced stars among the selected RGB stars,
we examined their [N/Fe] distribution as a function of [Fe/H], as shown in Figure \ref{fig2}.
The cyan error bar indicates the typical uncertainty of the measured [N/Fe] in the interval
of 0.5 dex in [Fe/H] from --2.0 to 0.0 and 1.0 dex from --3.0 to --2.0. They were derived using
the stars used to calibrate the method for determining [N/Fe] in
\citet{kim2022}. The typical error of the measured [N/Fe] appears
to increase with decreasing [Fe/H], and this
suggests that the measured error of [N/Fe] could affect the selection of N-rich stars,
especially at low metallicity. Thus, after incorporating
the uncertainty of the measured [N/Fe] as a function of [Fe/H],
we carried out a simple Monte Carlo (MC) simulation to derive
a selection function for the N-rich stars as follows.
Assuming a Gaussian error distribution, we retrieved [N/Fe] values
of individual stars using the error bar in the respective [Fe/H] bin in
Figure \ref{fig2}. We then calculated the mean and dispersion trends of [N/Fe]
as a function of [Fe/H], and defined as N-rich population (NRP) the stars that
are 2$\sigma$ or more away from the mean [N/Fe] trend at a given [Fe/H].
We performed this MC simulation 100 times, and by taking an average of
the 100 realizations we derived the mean and $2\sigma$ trend lines as
a function of [Fe/H]. However, one can guess that as the scatter of the RGB stars becomes
larger with decreasing metallicity, the 2$\sigma$ placement from
the mean also varies with the metallicity, that is
lower [N/Fe] at higher metallicity and larger [N/Fe] at lower metallicity. Because
other studies adopted the cut of [N/Fe] $>$ +0.5 to select N-rich
stars (e.g., \citealt{schiavon2017a,horta2021}), we
set the lower limit of the N-rich stars as [N/Fe] $\gtrsim$ +0.5 as well.
As a result, the region lower than [N/Fe] = +0.5 in the derived $2\sigma$ positions
was forced to set to [N/Fe] = +0.5. By doing so, we obtained the blue-solid
line as a selection function in Figure \ref{fig2}. We conservatively took this
line as a reference line to select N-rich stars in our study.
The selected N-rich stars are represented by red dots in
the figure, while the N-normal population (NNP) by the black dots.

We also verified from visual
inspection of their spectra that the [N/Fe] estimates of the selected
N-rich stars did not arise from defects
in their spectra. In addition, we cross-matched our giant sample with
those previously studied using the SEGUE, LAMOST, and APOGEE
data \citep[e.g.,][]{martell2011,martell2016,schiavon2017a,koch2019,tang2020,kisku2021,fernandez2022}.
We found 20 stars matched with the LAMOST giant sample studied by \citet{tang2020}, and they
turned out to be N-normal stars in our study. Twenty two stars
were found to overlap with the SEGUE giants studied by \citet{martell2011},
and 4 objects were classified as N-rich stars in our study.
We found 4 objects matched with the N-rich catalog of \citet{fernandez2022}, but unlike
their estimate, our derived [N/Fe] for them is slightly lower than
[N/Fe] $\sim$ +0.5, resulting in being classified as N-normal stars in our study.
The different classification among the studies is
caused by the selection method employed. \citet{tang2020} and \citet{martell2011}
used the line index of CN (at $\sim$ 3850 \AA) and CH (at $\sim$ 4300 \AA) features to identify
N-rich stars, which may not efficiently decouple degeneracy among the line
strength, temperature, and metallicity. Although \citet{fernandez2022} selected N-rich
objects based on high-resolution, near infrared spectra, we found slightly different
stellar parameters for four stars, which are blamed for their N-normal status.
In our study, we do not include them as N-rich stars for a self-consistent analysis.
As a result, a total number of 138 N-rich stars are
available for further analysis. Table \ref{tab1} provides the IDs of stars
that are classified as N-rich stars in our study.

\section{Space Velocity and Orbital Parameters} \label{sec3}

In order to characterize the dynamical properties of our N-rich stars, we derived their
space velocity components and orbital parameters. To compute the velocity components,
we adopted radial velocities from the SDSS and LAMOST spectra and proper motions
from Gaia Early Data Release 3 \citep[EDR3;][]{gaia2021}. We
did not attempt to estimate the radial velocity
from the SDSS and LAMOST spectra, as each survey reduction pipeline determines
the radial velocity of each star accurate to 5 -- 10 \kms\ depending on S/N,
which is the best achievable from low-resolution spectra \citep[e.g.,][]{yanny2009}.

Distances for the SDSS stars were determined following the methodology of \citet{beers2000,beers2012},
while distances for the LAMOST stars were obtained from the Value Added Catalog of
LAMOST DR5 \citep{xiang2019}. Moreover, to ensure that the two distance estimates
are on the same scale, we compared the distances of SDSS and LAMOST stars with
the ones determined using parallaxes from Gaia EDR3. Only stars with relative parallax errors
less than 10\% were retained; for these we removed the reported zero-point
offset of --0.017 mas \citep{lindegren2021}. We obtained
a difference of --0.031 mag in the distance modulus between the SDSS stars
and their Gaia parallax-based estimates, while for the LAMOST stars this offset was --0.001 mag We
adjusted the mean differences in our sample of stars to cancel out for these systematic offsets.

For the calculation of space velocity, we adopted $V_{\rm LSR} = 236$ \kms \citep{kawata2019}
for the rotation of the local standard of rest (LSR),
a solar position of \rsun\ = 8.2 kpc \citep{bland2016} from the Galactic center
and \zsun\ = 20.8 pc \citep{bennett2019} from the midplane. The adopted solar peculiar
motion with respect to the LSR is taken to be
$(U,V,W)_{\odot}$ = (--11.10,12.24,7.25) \kms \citep{schonrich2010},
where the velocity components $U$, $V$, and $W$ are positive in the directions toward the Galactic
anticenter, Galactic rotation, and North Galactic Pole, respectively. In addition, in a spherical
coordinate system with an origin at the Galactic center, we derived the velocity
components of $V_{\rm r}$, \vphi, and \vtheta.

We also computed various orbital parameters, such as maximum (\rmax)
and minimum (\rmin) distances from the Galactic center, maximum distance (\zmax)
from the Galactic plane, orbital eccentricity ($e$) defined by (\rmax\ -- \rmin)/(\rmax\ + \rmin),
and orbital inclination ($i$) by adopting an analytic
St$\ddot{a}$ckel-type potential, which has been employed by
numerous studies \citep{cb2000,carollo2007,carollo2010,kim2019,lee2019,kim2021,lee2023}.
We also derived the orbital elements of the MW's GCs by employing their positions,
proper motions, distances, and radial velocities, as provided by \citet{vasiliev2019}.

\begin{figure}
\centering
\includegraphics[width=0.95\columnwidth]{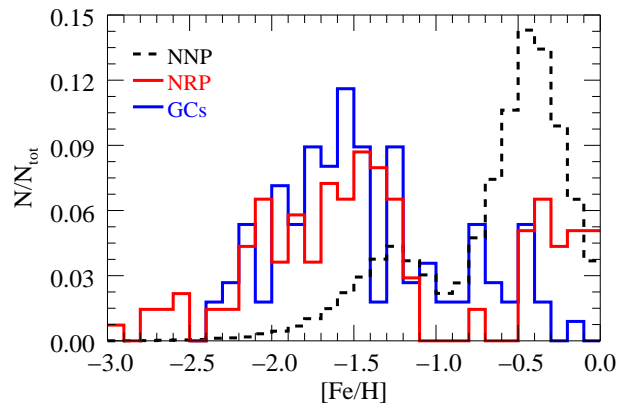}
\caption{Metallicity ([Fe/H]) distribution of NRP (red), NNP (black dashed), and
MW's GCs (blue). We did not include the GCs in the
Galactic bulge. The distribution is normalized by the total number of stars in
each population. The NRP is clearly more metal-poor than the NNP, and has an extended
distribution. The MDF of the GCs quite well overlaps with that of the NRP in the metallicity range
[Fe/H] $<$ --1.0. The metallicity values of the GCs were adopted from
the catalog by \citet{harris1996,harris2010}.}
\label{fig3}
\end{figure}

\begin{figure*}[t]
\centering
\includegraphics[width=0.9\textwidth]{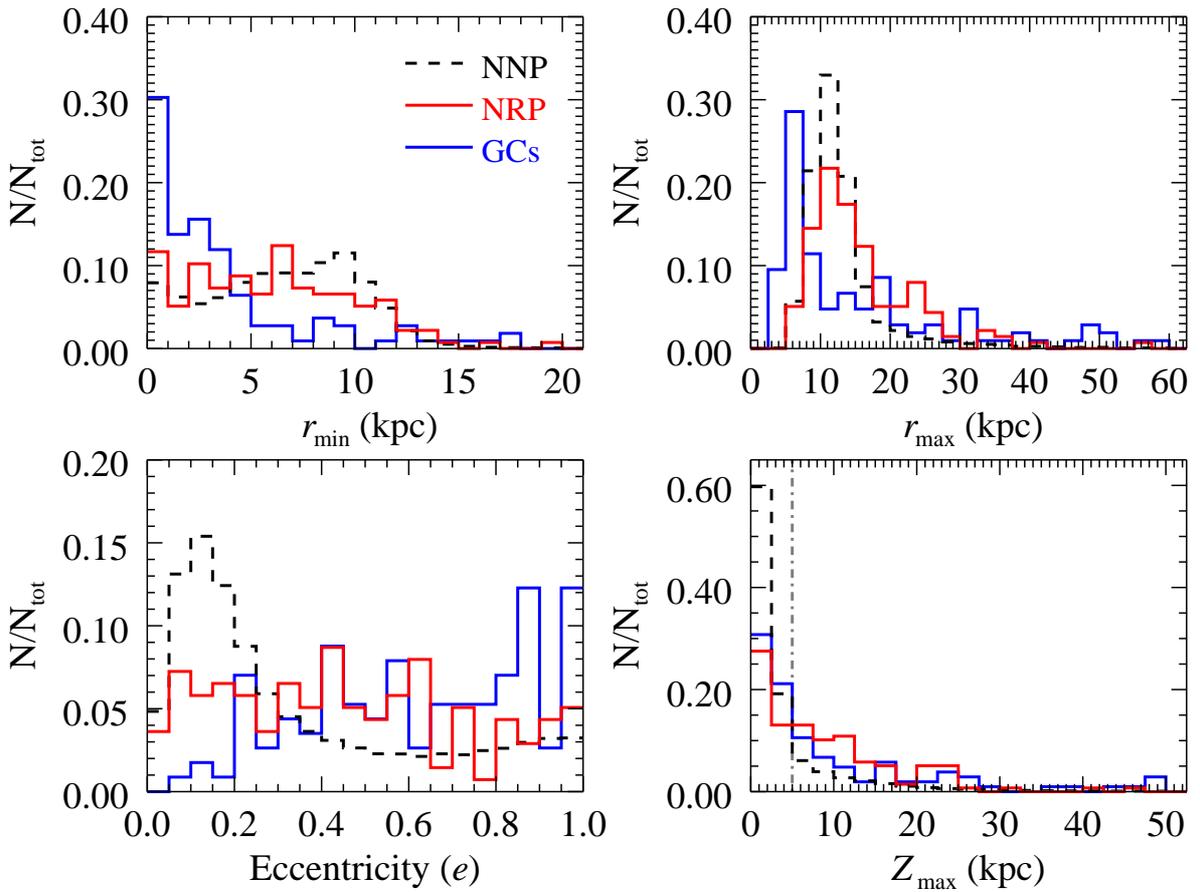}
\caption{Comparisons of orbital parameters (\rmin, \rmax, $e$, and \zmax) between
the NRP (red), NNP (black dashed), and GCs (blue). As
in Figure \ref{fig3}, we excluded the bulge GCs.
The distribution is normalized by the total number of stars in each group.
The gray vertical line in the bottom-right
panel delineates \zmax\ = 5 kpc, in which the Galactic halo system begins to dominate.}
\label{fig4}
\end{figure*}

\section{Results and Discussion} \label{res}

The observed chemical peculiarity of N-enhanced stars suggests that
they may have formed in environments different from those of N-normal stars.
One may expect that these different environmental origins should be reflected in their
metallicity distribution functions (MDFs) and dynamical properties.  In this section we explore
this question in detail, and consider its consequences.

\subsection{Contrast in the Metallicity Distribution Between NRP and NNP}\label{mdfcom}

We first examined the MDFs of our NNP and NRP, as shown in Figure \ref{fig3}.
The red histogram represents the NRP, whereas the black one applies to the NNP.
We also plotted the MDF of the MW's GCs for comparison. Note that
we excluded the bulge GCs, which are identified in \citet{massari2019},
because our NRP does not include the stars near the Galactic bulge.
The figure shows that the MDFs
of both the NRP and NNP have two peaks, well-separated around [Fe/H] $\sim$ --1.0, with
some discrepancies between them. The NNP is represented by well-defined MDFs of the
metal-rich disk and metal-poor halo populations. The disk population is dominated by the
thick-disk stars and the Splash \citep{belokurov2020} with [Fe/H] $>$ --1.0,
while the (inner-) halo population is heavily influenced by GSE with [Fe/H] $<$ --1.0.

The NRP has a metallicity distribution biased toward the
more metal-poor and metal-rich regions; the MDF of the metal-poor stars ([Fe/H] $<$ --1.0)
has an extended distribution down to [Fe/H] = --3.0, while the metal-rich
stars ([Fe/H] $>$ --1.0) cut off at [Fe/H] $\sim$ --0.5. One clear discrepancy
is that the NRP is dominated by more metal-poor stars
with [Fe/H] $<$ --1.5 than the NNP. This chemical
dissimilarity between the NNP and NRP suggests that the NRP may have
gone through a different chemical-evolution history than the NNP. A Kolmogorov-Smirnov (K-S)
two-sample test indicates that the probability that the two groups of stars are
drawn from a same parent population is lower than 0.1\%, clear
evidence for different origins.

Figure \ref{fig3} also reveals that the MDF of the NRP mimics that of the MW's GCs
for [Fe/H] $<$ --1.0. A K-S
test of the MDFs in this metallicity range yields a $p$-value of 0.92,
in other words, that the distributions are indistinguishable, implying
that a large fraction of the NRP may be escapees from existing GCs within
this metallicity range. On the other hand, we obtained a $p$-value less than 0.001 in the
same metallicity range between the GCs and NNP, a clear signature
that the NNP does not come from existing GCs. The rather different shape of the MDF
for [Fe/H] $>$ --1.0 between the NRP and GCs and the wide metallicity range of the NRP
suggest diverse birth places for stars in the NRP.

\subsection{Contrast in Dynamical Characteristics Between NRP and NNP}

We further searched for distinguishable properties between the NNP and NRP in their
respective orbital parameter spaces, as shown in Figure \ref{fig4}.
The typical uncertainty of each parameter is 0.6, 1.6, 1.4 kpc,
and 0.04 for \rmin, \rmax, \zmax, and $e$, respectively.
In the figure, the NRP is represented by the red histogram, while the NNR is the black-dashed one.
In each panel, we also included the distributions of the GCs for
comparison. The histograms are normalized by the total number of stars in each population.
A number of distinct trends are apparent. The top-left panel indicates that
the \rmin\ distribution of the NNP peaks at \rmin\ $\sim$ 9 kpc, whereas that of
the NRP gradually declines with increasing \rmin. The \rmax\
distribution for both populations peaks at \rmax\ $\sim$ 11 kpc, but there is a
lingering trend of the N-enhanced stars found at larger
\rmax\ than the N-normal ones. Because both the NRP and NNP have \rmax\ $>$ 5.0 kpc, no
stars are bound to the Galactic bulge.

These different behaviors between the two populations are directly
reflected in the eccentricity ($e$) distribution shown in the bottom-left panel.
The NNP is mostly populated within $e \lesssim 0.4$, which is associated with the disk system,
while the NRP is distributed more evenly, leading to higher fraction of high-$e$ stars
compared to the NNP.

In the case of \zmax, as seen in the bottom-right panel, more fraction of stars in the
NRP exhibit excursions that reach farther away from the Galactic plane than the NNP and
the GCs. The gray-dashed line is located at \zmax\ = 5 kpc, beyond which Galactic halo
stars begin to dominate. Judging from this reference line, the majority of the NRP
travel through the Galactic halo region, whereas most of the NNP are confined to the
Galactic disk system. Application of K-S two-sample tests for these orbital parameters
yielded $p$-values smaller than 0.001 for all of them, under the null hypothesis that
the two groups of stars share the same parent population.
These disparate dynamical properties provide further evidence
for the distinct origins of the NRP and NNP.

Contrasting dynamical properties between the NRP and NNP
have been reported in other studies. For example, \citet{tang2020}
carried out a similar kinematic analysis using about 100 metal-poor CN-strong stars
from LAMOST DR5. Although their RGB sample covers a much narrower metallicity
range --1.8 $<$ [Fe/H] $<$ --1.0 than ours (--3.0 $<$ [Fe/H] $<$ 0.0), they came to a
similar conclusion, that their N-enhanced and N-normal stars do not
share the same birthplace. Based on the finding that the orbital properties of
their N-rich field stars are very similar to MW inner-halo stars, they concluded
that their N-enhanced stars are likely to have originated from GCs.

We also compared the dynamical properties of the NRP with
those of the GCs (blue histogram in Figure 4).
Because we found very similar MDFs in the region of [Fe/H] $<$ --1.0
between the NRP and GCs in Figure \ref{fig3},
we carried out K-S two-sample tests in the same metallicity region for the four
orbital parameters. Interestingly, we obtained $p$-values lower than 0.001
for \rmin, \rmax, and $e$, while \zmax\ has a $p$-value of 0.06.
Given these low $p$-values, which imply different dynamical properties of their
parent populations, our NRP does not appear to originate in the current MW's GCs,
which contradicts the interpretation from the MDF comparison.
For these disparate characteristics, one may naively think of the contribution from fully
dissolved GCs, which had metallicities similar to those of the MW's GCs,
resulting in the similar MDFs, but dissimilar dynamical properties.
This may be partially true for the stars with [Fe/H] $<$ --2.5, because there
are no MW's GCs with [Fe/H] $<$ --2.5. However, a more plausible
explanation for the dynamical discrepancies is that some of the NRP stars in
the metallicity range [Fe/H] $<$ --1.0 may be indeed escaped from GCs in the past, and
they have dissimilar properties with the current GCs, because the dynamical properties of the GCs
have been changed by the dynamic friction of the MW over time, while the escaped stars
have not. From the comparison between NNP and GCs, we obtained
the $p$-values of all parameters lower than 0.001, implying that the NNP is not
associated with the GCs, which is consistent with the MDF comparison.

Our interpretation above partially agrees with other studies.
For example, \citet{savino2019} compared the orbital
properties of 57 CN-strong field stars observed in SDSS with those of the MW's GCs.
They found that only 20 stars (35\%) have orbital characteristics similar
to the existing GCs, and recognized that the CN-strong stars with halo
kinematics are expected under the GC-escapee scenario. \citet{carollo2013} also reached
a similar conclusion, based on the fact that both their N-enhanced field stars and
the MW GCs exhibit inner-halo kinematics and orbital properties.
In order to confirm which scenario is more realistic, it is
necessary to carry out a more detailed chemical-abundance
analysis of the NRP stars, and compare their abundance patterns
with those of the existing GCs.

\begin{figure}
\centering
\includegraphics[width=0.93\columnwidth]{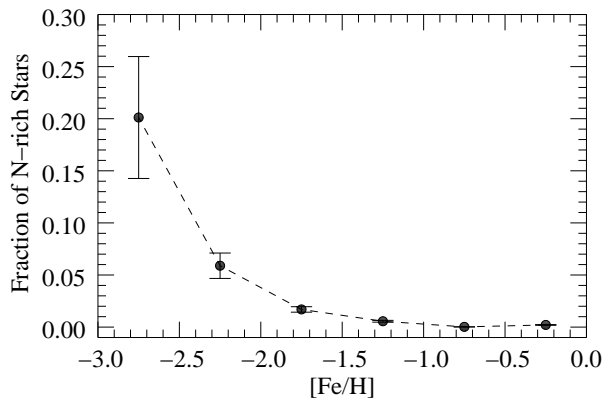}
\caption{Frequency of N-rich stars as a function of [Fe/H]. The error bars
are derived by bootstrap resampling.}
\label{fig5}
\end{figure}

\subsection{Fractions of N-rich Stars}

Given that the N-rich stars originated in GCs,
the fraction of N-rich stars as a function of [Fe/H] can provide an important
clue to characterizing their host systems, as well as the level of
their contribution to the Galactic halo.
We begin by plotting the frequency of the N-rich stars with
respect to the halo field stars as a function
of [Fe/H], as shown in Figure \ref{fig5}. The metallicity bin size to calculate the fraction is 0.5 dex,
and the error bars are based on bootstrap resampling of our programs stars.
Inspection of this figure reveals that the N-rich fraction
does not increase much for [Fe/H] $>$ --2.0, while
a significant rise of up to $\sim$20\% occurs for [Fe/H] $<$ --2.0,
indicating a larger contribution of the GC SG stars with [Fe/H] $<$ --2.0.
The fact that there are not many GCs with [Fe/H] $<$ --2.0 (and no intact
GC with [Fe/H] $<$ --2.5) leads us to conjecture that the fully destroyed
very metal-poor (VMP; [Fe/H] $<$ --2.0) GCs may have
in some degree contributed to the Galactic halo.
As the MW's metal-poor GCs are generally older than the metal-rich counterparts
\citep{massari2019}, they have experienced more prolonged tidal interactions with the MW and
were fully destroyed.

If this is the case, it may have an interesting impact on the frequency of the
carbon-enhanced metal-poor (CEMP; [Fe/H] $<$ --1.0, [C/Fe] $>$ +0.7)
stars in the Galactic halo. Studies show that the fraction of the CEMP stars
accounts for $\sim$ 10 -- 20\% between [Fe/H] = --3.0 and --2.0 \citep{lee2013,placco2014}.
One can expect that, without the contribution of FG and SG stars
from partially or fully dissolved GCs, which are mostly
carbon-normal stars, the CEMP frequency in the metallicity range
would be much higher. The observed fraction of CEMP stars has been frequently used to
constrain the initial mass function (IMF) of the MW using
population-synthesis models (e.g., \citealt{komiya2007,suda2013,lee2014}).
The model predictions of the CEMP frequency are somewhat higher than
the observed trend in that metallicity range. By taking into account the contribution
of the GC stars to the MW, the deficit in the CEMP frequency between the model and observation
can be reduced, and in turn it will provide more strict constraints to the models for
prediction of the IMF.

The overall NRP fraction allows us to estimate the GC
contribution to the mass budget of
the MW, and a number of studies have attempted to predict this over the past few years.
For instance, using the metal-poor halo giants (--1.8 $<$ [Fe/H] $<$ --1.0)
from APOGEE DR12, \citet{martell2016} attempted to calculate the fraction of N-rich stars
with respect to the halo field stars, deriving a contribution of
2 -- 3\% of GC SG stars in the halo. Their
previous work also accounted for the similar percentage \citep{martell2011}.
This has been further confirmed by \citet{koch2019}, who used CN strong stars from
SEGUE data to compute about 2.6\%. \citet{horta2021} also estimated a fraction of 2.7\%
at 10 kpc from the Galactic center, using the APOGEE DR16 \citep{jonsson2020} giant sample.
We derived a fraction of 0.4 $\pm$ 0.1\% of the N-rich stars in our RGB sample,
which is about 6 to 7 times lower than those studies. Even if we restrict ourselves
to the metallicity range --2.0 $\leq$ [Fe/H] $\leq$ --1.0, on which the
previous studies focused, we obtained a fraction of 0.8 $\pm$ 0.1\%.
The main reason for the much lower value is that our NNP sample is overwhelmed by
local halo stars dominated by the GSE structure, with a MDF peak at [Fe/H] $\sim$ --1.3,
as can be appreciated in Figure \ref{fig3}. These stars
were mostly observed in the LAMOST survey, and they are
located closer to the Galactic disk than the SDSS stars. Consequently, our sample includes
more N-normal disk stars than the previous studies, as can be inferred from
the bottom panels of Figure \ref{fig4}. In this respect,
considering only the SDSS giants, which reside in more distant halo region, we
obtained a fraction of 2.3 $\pm$ 0.2 \% for the N-rich stars,
which well agrees with those of \citet{martell2011} and \citet{koch2019}, who utilized
the SEGUE spectra to derive the fractions. The different temperature range
for selecting giants could be another source of the large discrepancy. The
previous studies mentioned above include more cool giants (\teff\ $<$ 4500 K), in contrast
to our giant sample.

We note that the observed N-rich fractions do not directly reflect the contribution
of the GC-origin stars to the Galactic halo. Depending on the assumed ratio
between the FG and SG of a GC and the dissolved fraction of a GC, the total
GC contribution to the halo of the MW significantly varies from study to
study (17 -- 50\%) \citep{martell2010,martell2011,carollo2013,koch2019,tang2020,horta2021}.
As those parameters for GCs are not well-established at present,
in this study we do not attempt to evaluate the total fraction of the GC-origin stars
using the derived fraction of the N-rich stars.

Even though there are no bulge stars in our sample, it is worthwhile mentioning
the situation for the NRP faction in the Galactic bulge, as it is quite different
from the Galactic halo. \citet{schiavon2017a}, who focused on stellar populations
within $\sim$ 3 kpc of the Galactic center, derived a fraction of 13 -- 17 \% of
N-rich stars, which is much higher than found in the halo region.
Using giants from APOGEE DR16 and a density model
of the halo stellar population, \citet{horta2021} estimated a fraction of
16.8\% of N-rich stars with respect to N-normal stars at a distance of 1.5 kpc
from the Galactic center. They further predicted, by assuming
the ratio of 0.5 between the FG and SG in a GC proposed by \citet{schiavon2017a}
that the total contribution of disrupted GC stars to the MW accounts
for 27.5\% and 4.2\% at 1.5 kpc and 10 kpc from the Galactic center, respectively.
As discussed in \citet{horta2021}, the higher fraction of the N-rich stars in the inner
region of the MW may be explained by more frequent merger events in the early history of
the MW. During such mergers, some fraction of extragalactic-origin N-rich
stars were brought into that region. In addition, the in situ formation and destruction
of GCs were actively contributing, resulting in a large number of N-rich stars in the inner
region of the MW. Overall, the observational data suggest that the
disrupted GCs donated their stars
to both the halo and bulge, with greater contributions to the bulge.

\begin{figure*}[!t]
\centering
\includegraphics[width=0.95\columnwidth]{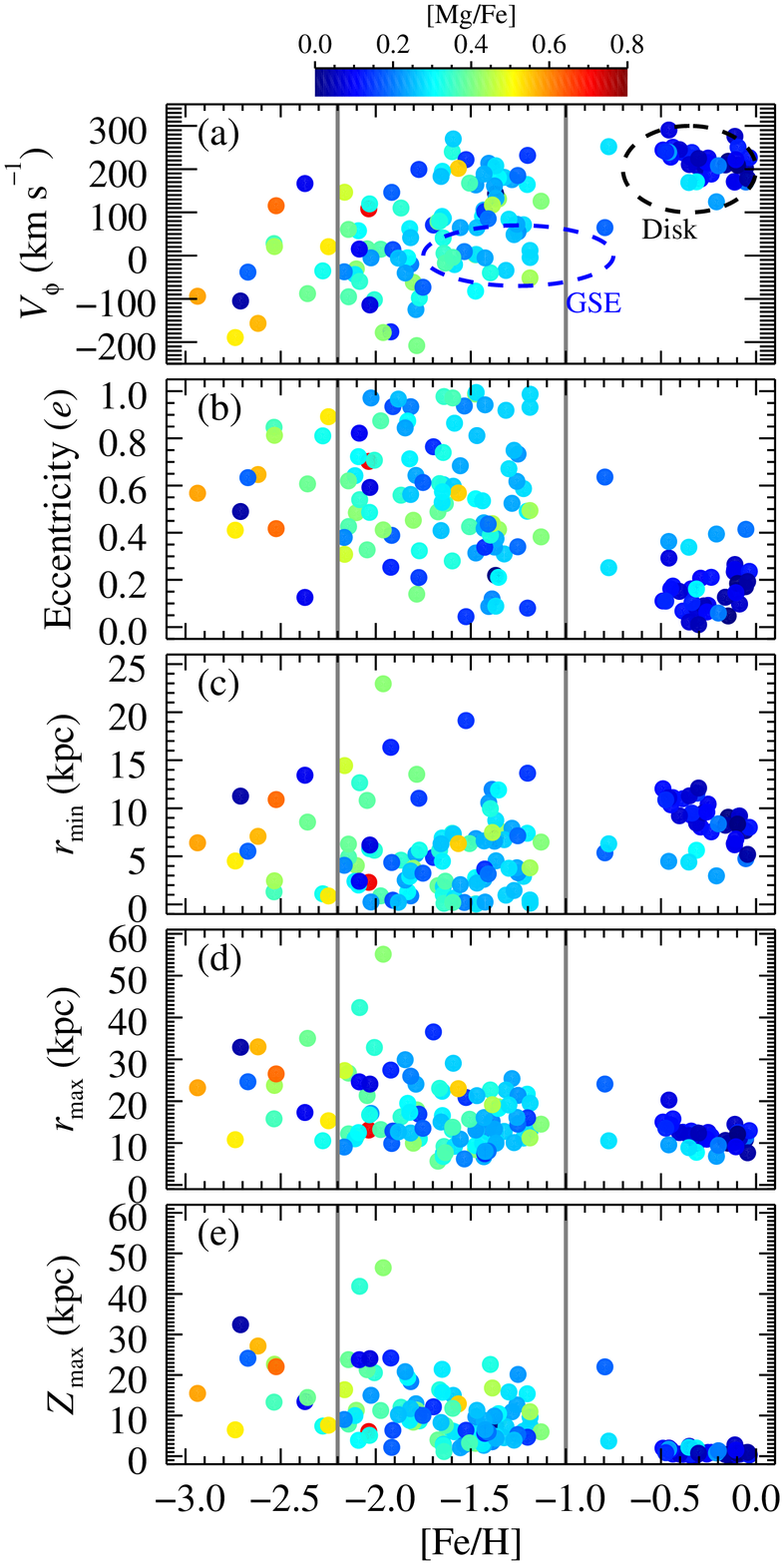}
\includegraphics[width=0.95\columnwidth]{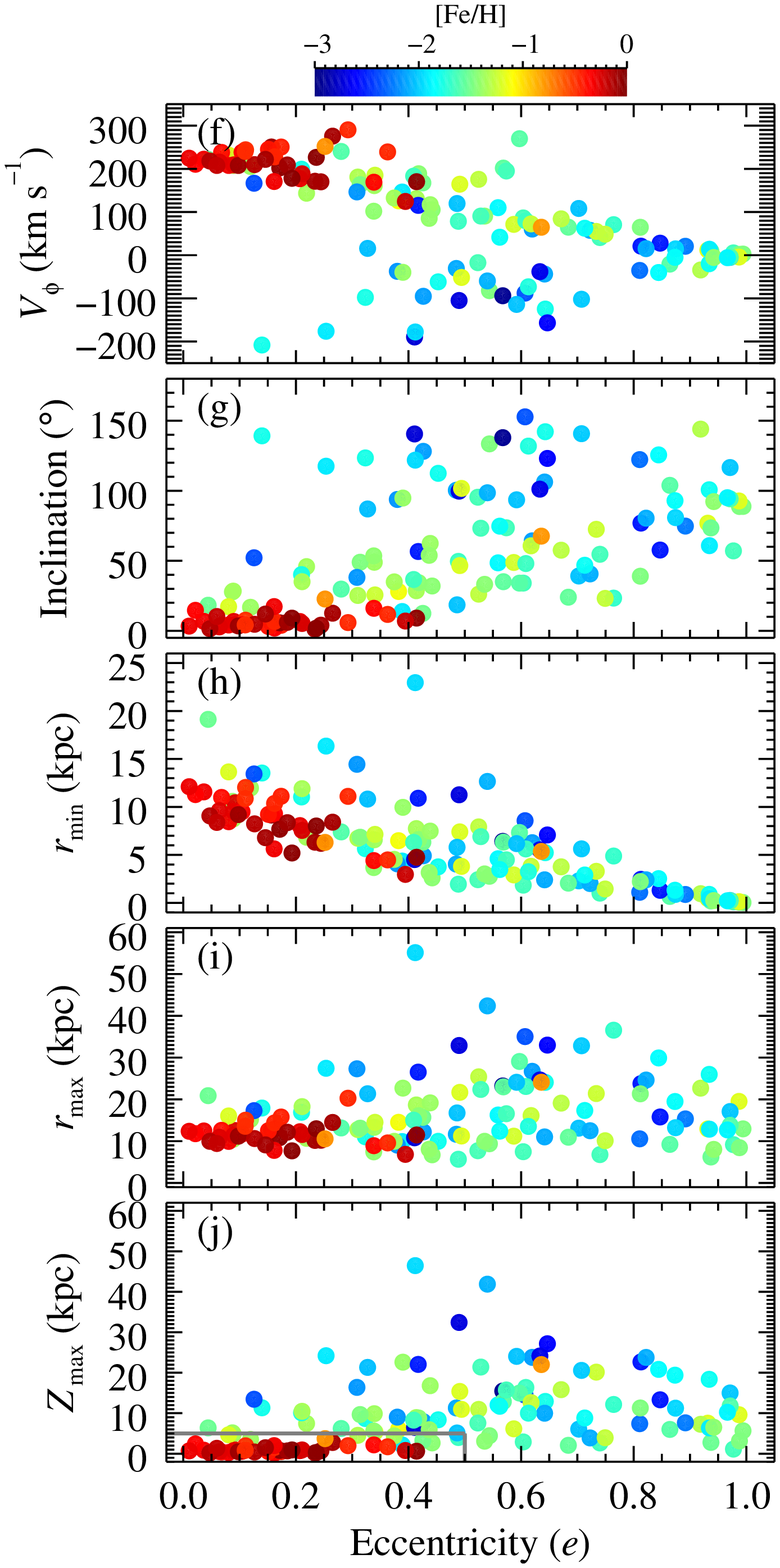}
\caption{Left panels: Runs of \vphi, eccentricity ($e$), \rmin, \rmax, and \zmax\
with respect to [Fe/H] for stars in the NRP. The color code represents [Mg/Fe], shown
in the color bar above the column of panels. In Panel (a),
the black-dashed circle represents the Galactic disk system, while the blue-dashed ellipse
is the approximate region of the GSE structure based on \citet{belokurov2018}.
The vertical lines denote to separate the NRP into metal-rich ([Fe/H] $>$ --1.0),
N-rich (MRNR), metal-intermediate (--2.2 $<$ [Fe/H] $\leq$ --1.0), N-rich (MINR), and
metal-poor ([Fe/H] $\leq$ --2.2), N-rich (MPNR) stars. Right panels: Runs of \vphi,
inclination ($i$), \rmin, \rmax, and \zmax\ with respect to eccentricity
for stars in the NRP. The color code represents [Fe/H], shown in the color bar above
the column of panels. The gray box in Panel (j) indicates the division for the in situ
origin adopted by \citet{massari2019}. See Section \ref{inex} for a detailed description.}
\label{fig6}
\end{figure*}

\subsection{Detailed Dynamical Properties of the NRP}

It has been recognized for more than a decade that the Galactic GCs comprise both those that
formed in situ and those that were accreted
(e.g., \citealt{forbes2010,massari2019,myeong2019,vasiliev2019}). This
means that the stars in the NRP may also originate from in situ formed GCs or accreted GCs.
A detailed dynamical analysis of the NRP may be able to distinguish the
progenitors of the stars in the NRP into the in situ and accreted components.

Several studies have been carried out in this regard. \citet{kisku2021}
found from APOGEE DR16 that, among
their N-rich stars located near the Galactic bulge, the accreted stars
account for about 30\% of the total. They separated the accreted
component using the Si abundance ratio in the [Si/Fe] versus [Fe/H] plane, and
kinematically confirmed the separation as well.
\citet{tang2020} also cautiously reported, based on a small number of stars (4 out of 11 stars)
that a substantial portion (36\%) of their N-rich field stars may have been accreted by the MW.
From a more extensive study, \citet{fernandez2022} compiled a total of 412 unique N-rich
stars that are likely to have been stripped from GCs, and
they demonstrated through a kinematic analysis that they must have originated
from several channels, including dissolved members of GCs
brought into the MW during major-merger events such as the GSE.

Here, in order to investigate their dynamical properties and qualitatively assess the origin
of their progenitors, we examine the locations of our N-rich stars in various orbital
parameter spaces, as a function of [Fe/H] (left) and orbital eccentricity (right), as shown in
Figure~\ref{fig6}. The color code indicates [Mg/Fe] (left panels) and [Fe/H] (right panels) scales,
and each scale is represented by the color bar at the top. In Panel (a),
the black-dashed circle represents the Galactic disk system, while the blue-dashed ellipse
is the approximate region of the GSE structure based on \citet{belokurov2018}.

The overall behavior of the NRP in Figure~\ref{fig6} is as follows.
The rotation velocity \vphi\ exhibits a good correlation with [Fe/H] (Panel (a)).
The orbital parameters of \rmax\ and \zmax\ show anti-correlations with [Fe/H] for
[Fe/H] $<$ --1.0. The anti-correlation becomes more
prominent below [Fe/H] = --2.2, which is the peak of the MDF of the outer-halo
stars (e.g., \citealt{carollo2010,beers2012,lee2017}).
We also note a well-established anti-correlation between \rmin\ and $e$ in Panel (h).
Upon close inspection of the left-column panels, we realized that
it is convenient for our analysis to divide our NRP stars
into at least three different groups according to the metallicity:
metal-rich ([Fe/H] $>$ --1.0), N-rich (MRNR),
metal-intermediate (--2.2 $<$ [Fe/H] $\leq$ --1.0), N-rich (MINR), and
metal-poor ([Fe/H] $\leq$ --2.2), N-rich (MPNR). The two
vertical lines delineate the domains of this separation. We can
confirm that this chemical separation is qualitatively justified by well-behaved
distinct dynamical properties of each metallicity group in the right-column panels,
and discuss these groupings in detail below.

\subsubsection{Metal-poor, N-rich (MPNR) Group}

The majority of the MPNR stars possess high [Mg/Fe] ($>$ +0.3), high
eccentricity ($e >$ 0.4), and high inclination ($i >$ 60$^{\circ}$).
Note that stars with $i >$ 90$^{\circ}$ have retrograde motions.
We also note that most of the MPNR stars exhibit relatively
low \vphi\ ($<$ 50 \kms) and even retrograde motion, and some of them
travel farther out up to \rmax\ $\sim$ 35 kpc
and \zmax\ $\sim$ 30 kpc into the halo. Interestingly, a few stars even approach
close to the Galactic bulge (\rmin\ $<$ 3 kpc).
These dynamical characteristics may indicate that the host systems of some of the
MPNR stars are accreted; in particular all stars with [Fe/H] $<$ --2.5 would have an ex situ
origin, as no surviving MW GCs have such low metallicity. Their low \vphi\ ($<$ 50 \kms) also
provides more evidence for an accretion origin. The hypothesis of extragalactic origin is supported
by a recent study, which claimed that
some of the N-rich stars in the Galactic halo are accreted from the GCs associated
with Sagittarius (Sgr) dwarf galaxy \citep{fernandez2021a}. The theoretical work by \citet{rostami2022} also
upholds this idea. These authors simulated the escaped fraction of GCs in a dwarf galaxy,
and argued that many dwarf galaxies, such as Sgr, Fornax, Sculptor, etc.
in the MW have lost some of their GCs while interacting with the MW.

One may think that the stars with moderate rotation velocity (\vphi\ $>$ 50 \kms) and
[Fe/H] $>$ --2.5 originate from partially or fully disrupted GCs, which formed in situ because
the MW's GCs exhibit an average rotation velocity
of \vphi\ = 50 -- 80 \kms\ \citep{vasiliev2019}.

\subsubsection{Metal-intermediate, N-rich (MINR) Group}

The MINR stars have a wide range of \vphi, $e$, \rmin, and $i$, while
their \rmax\ and \zmax\ values are concentrated in a relatively narrow
range. These features imply diverse origins at similar metallicities.

According to \citet{vasiliev2019}, the GCs within 10 kpc from the
Galactic center have a mean azimuthal velocity
in the range 50 -- 80 \kms, while the ones in the outer region exhibit
strongly radial orbits. In addition, the GCs that reside in the inner
halo are relatively more metal-rich ([Fe/H] $>$ --2.0). These aspects suggest that
a large fraction of our MINR stars with moderate rotation velocity (\vphi\ $>$ 50 \kms)
and low $e$ ($<$ 0.4) may have been stripped from existing GCs, which is in accordance
with the interpretation from the MDF comparison in Section \ref{mdfcom}.

Among our MINR stars, the ones in the blue ellipse in Panel (a)
with high eccentricity ($e >$ 0.7) and \vphi\ $\sim$ 0.0 \kms\
may have originated from disrupted GCs that belonged to the GSE during its merger
with the MW, as their dynamical properties coincide with those of the GSE.
Indeed, \citet{myeong2018} reported about 8 GCs associated with the GSE, with
$e$ $>$ 0.8 and \vphi\ $<$ 50 \kms. We
observe that a small fraction of our MINR stars have orbits that pass close to
the Galactic center (\rmin\ $<$ 3 kpc, see Panel (h)),
but are not bound to the bulge, because of their \rmax\ $>$ 5 kpc.
Since these stars have strong radially biased orbits and $e >$ 0.8,
their host GCs are highly likely to originate from the GSE merger event,
as claimed by \citet{myeong2018}. The fact that
the GSE stars with [Fe/H] $<$ --1.0 on average
exhibit [Mg/Fe] $>$ +0.1 (e.g., \citealt{lee2023}) reinforces
the above claim. Note that, of the likely members of the GC SG, the stars
with relatively low Mg abundance ratios ([Mg/Fe] $<$ 0.0) are regarded as
an extragalactic origin, because the low Mg abundance ratio along with high Al is reported
in some GCs in M31 and the Large Magellanic
Cloud (\citealt{colucci2009,colucci2012,fernandez2020b}).
Other studies also reported low-[Mg/Fe] stars among the candidates for dissolved GC
members \citep{fernandez2016,fernandez2017,fernandez2019}. Thus, we expect
that at least a  few stars with [Mg/Fe] $\sim$ 0.0 among our MINR stars may fall in that category.

We also note a handful of stars with orbital eccentricities $e$ = 0.4 -- 0.6 and chemical
abundances of [Fe/H] $\sim$ --1.3 and [Mg/Fe] $\sim$ +0.2, which are
characteristic of the Sgr dwarf galaxy \citep{fernandez2021a}. As \citet{fernandez2021a}
reported possible member stars of M54, these stars are likely to have come from
disrupted GCs of the Sgr dwarf galaxy. Furthermore, because a small portion of the MINR stars share
metallicity ([Fe/H] $<$ --1.7) and dynamical properties (\vphi\ $<$ --50 \kms\ and $e$ = 0.5 -- 0.6)
with those of the Sequoia event \citep{myeong2019,naidu2020}, we cannot rule out their association
with the Sequoia dwarf galaxy. Through a kinematic analysis of MW's GCs,
\citet{massari2019} also claimed that about 5\% of the MW's GCs have originated from the
Sequoia event.

\subsubsection{Metal-rich, N-rich (MRNR) Group}

The MRNR population ([Fe/H] $>$ --1.0) has mostly low eccentricity ($e \lesssim$ 0.3) and
low inclination ($i \lesssim$ 20$^{\circ}$) with high prograde motions (\vphi\ $\gtrsim$ 160 \kms)
and low \zmax\ ($\lesssim$ 3 kpc), indicating that they are
confined close to the Galactic plane. We also observe that most of the MRNR stars are
located in the rotationally supported inner-disk region (\rmin\ $<$ 12 kpc), and do not
travel far away (\rmax\ $<$ 16 kpc) from the Galactic center. Their
Mg abundance ratio is low ([Mg/Fe] $<$ 0.2), as in disk stars.
Because our MRNR stars do not
follow the MDF of the existing GCs (see Figure \ref{fig3}), and their MDF is
shifted to the more metal-rich region, they are not likely to be
connected with the existing GCs. Rather, these dynamical and chemical features suggest that
most of the MRNR stars may have originated from fully destroyed metal-rich GCs formed
in situ in the disk, as envisaged by \citet{fernandez2021b}.

However, there are a few stars with intriguing distinct kinematics
and somewhat high [Mg/Fe] $\gtrsim$ +0.2, which diverge from the typical
trend. Among them, one star draws our attention. Its eccentricity and metallicity
are $e \sim$ 0.6 and [Fe/H] $\sim$ --0.8. This star may be associated with the progenitor of the GSE
merger event, because it is located near the GSE structure in Panel (a) of Figure \ref{fig6}.
It is also plausible that it originates
from the GSE-induced starburst, as recently reported by a few studies \citep{myeong2022,an2023,lee2023}.
The other star, at [Fe/H] $\sim$ --0.8 and \vphi\ $\sim$ 240 \kms, may be in this category
as well. However, it cannot be ruled out that, because its metallicity and eccentricity are
overlapped with the Splash stars, it could be a dynamically heated star from the GSE merger
after its host GC formed in situ and was disrupted in the disk. Its relatively high [Mg/Fe]
value ([Mg/Fe] $\sim$ +0.2) supports this scenario. It is known that the GSE stars
exhibit low [Mg/Fe] ($<$ +0.1) at [Fe/H] $>$ --1.0, while the Splash shows
high [Mg/Fe] ($>$ +0.2) for [Fe/H] $>$ --1.0 \citep{belokurov2020,lee2023}. A more detailed
chemical-abundance pattern is required to unveil this star's origin.

\subsubsection{In Situ Versus Accretion}\label{inex}

Taking into account our qualitative assessment of the possible origins of our NRP stars
from the dynamical characterization, it is clear that most of our NRP stemmed from diverse
groups of in situ formed GCs and ex situ formed GCs. There may not
be clear cuts to separate the accreted origin from our NRP sample. Nonetheless,
based on the chemical and dynamical properties of the NRP, at least
we can follow the procedure devised by \citet{massari2019} to identify
the stars with the accretion origin. They carried out a kinematic analysis of MW's GCs with their ages,
and divided them to in situ and accreted populations. They found that 40\% of the GCs
likely formed in situ, and 35\% are possibly associated with GSE, the Sgr dwarf galaxy,
the progenitor of the Helmi streams, and the Sequoia galaxy. The rest are likely
to have heterogeneous origins.

By simply adopting their approach, we can divide our NRP stars into in situ and
accreted components, as shown in Panel (j) of Figure \ref{fig6}. Note that in the panel,
instead of the circularity that \citet{massari2019} used, we employed the eccentricity.
Both parameters are good indicators for the orbital paths of a star.
We regarded the N-rich stars with \zmax\ $<$ 5 kpc, $e$ $<$ 0.5, and
\vphi\ $>$ 0 \kms\ as the in situ population. We adopted $e$ $<$ 0.5,
since the disk stars can have the eccentricity as high as $e$ $\sim$ 0.5 \citep{lee2011b,han2020}.
These stars mostly correspond to the ones inside the gray box in the panel.
From this exercise, we obtained a fraction of 33.5 $\pm$ 4.7\% for the in situ component,
which is not far from that of \citet{massari2019}.
We note, however, that by comparing with the E-MOSAICS simulations of MW-mass galaxies
and their GC populations \citep{pfeffer2018,kruijssen2019a}, \citet{kruijssen2019b}
predicted that 67 (43\%) out of 157 MW's GCs in the \citet{harris1996,harris2010}
catalog were accreted, resulting in a fraction
of 57\% for the in situ origin, which is higher than ours and that of
\citet{massari2019}.

In addition to the above exercise, we have carried out simple calculations
to estimate the fraction of the stars with accretion origin in each metallicity
group by distinguishing them from those with an in situ origin. For this to work,
we introduced three criteria for the in situ origin: \vphi\ $>$ 50 \kms, \zmax\ $<$ 5 kpc,
and $e$ $<$ 0.7.  The rotation velocity cut comes from the fact that the overall rotation
velocity of the GCs within 10 kpc from the Galactic center is larger
than 50 \kms\ \citep{vasiliev2019}.
It is also taken into account that the GSE stars have on average \vphi\ = 0 \kms,
with a \vphi\ dispersion of $\sim$ 50 \kms\ \citep{belokurov2020}.
The \zmax\ condition is adopted from \citet{massari2019}. The last criterion is reflected
by eliminating the GSE-origin stars, as most of them have high $e$ ($>$ 0.7).
We imposed the $e$ cut to only the MINR group, because it has many stars
with GSE kinematics. We regarded the stars that do not simultaneous satisfy
these cuts as having an ex situ origin.

By applying the above cuts to the MPNR, MINR, and MRNR groups, we obtained
an accreted fraction of 98 $\pm$ 2\% for the MPNR group, 86 $\pm$ 4\% for the MINR group,
and 3 $\pm$ 3\% for the MPNR group; the total fraction of stars with ex situ origins among
our NRP is 64 $\pm$ 4\%. The fraction and its associated
error were the mean and standard deviation derived by a Monte Carlo simulation of
resampling each group of stars 10,000 times. When resampling, the uncertainty of
each parameter was incorporated. Of course, all of the
stars in these fractions may not have an ex situ origin, as pointed
out by \citet{massari2019}, and our derived fraction may only suggest that
a substantial number of the N-rich stars may have the ex sit origin.

Our derived fraction (64\%) of the NRP stars with an ex situ origin is
much higher than 43\% reported by  \citet{kruijssen2019b}. This is
mainly due to how we define the ex situ origin, especially using
the dynamical properties. As \citet{kruijssen2019a} demonstrated
from their simulations of the E-MOSAICS galaxies,
about half of the GCs with [Fe/H] $<$ --1.5 are predicted to have
formed in situ in the early assembly history of the MW.
Providing that these metal-poor GCs do not exhibit disk-like kinematics
at present, which is highly likely because they
may have experienced continuous perturbations by the MW over
a protracted period, they tend to be assigned an ex situ origin, resulting in
a higher fraction of the ex situ NRP stars. Considering this, and the fact
that the current GCs of the MW are mostly located within \zmax\ = 10 kpc, as can
be seen in the bottom-right panel of Figure \ref{fig4}, if we relax
the condition for the in situ origin to have \zmax\ $<$ 10 kpc, we obtain a slightly
lower fraction of 53\% for the ex situ origins. Interestingly,
using the Na-abundance ratios measured from the SDSS low-resolution spectra,
\citet{koo2022} derived a fraction of 67.8\% for the accreted
origin among the Na-enhanced stars, which are believed to originate from GCs,
in excellent agreement with our estimate (64 $\pm$ 4\%) within the error.

To sum up, the existence of dynamical distinctions in each
chemical group among our NRP stars indicates that the stars within each group
do not share a single common origin. A simple method for
distinguishing the in situ origin reveals that, although
about 34\% of our NRP stars seem to share an origin with the in situ
formed GCs, the origins for the rest are not associated with those
of typical Galactic populations, but are of extragalactic origin, likely to be
associated with the GSE, Sgr dwarf galaxy, and Sequoia.
However, as we have emphasized, the fraction of accreted origin stars strongly depends
on how they are defined.

\section{Summary and Conclusions} \label{sum}

We have identified 138 N-rich stars from low-resolution SDSS and LAMOST spectra,
covering the metallicity range of --3.0 $<$ [Fe/H] $<$ 0.0, and compared
their chemical and dynamical properties with those of the N-normal stars.
The MDF of the NNP is well-matched by the canonical metal-rich disk
and metal-poor halo populations, whereas the NRP has a more
extended distribution toward the metal-rich and metal-poor regime.
The MDF of the NRP looks very similar to that of the MW's GCs
in the range of [Fe/H] $<$ --1.0, but,
K-S tests for the orbital parameters yielded $p$-values of less than
0.001 in the same metallicity range. Theses results may suggest that
the escaped stars from the existing GCs account for a very small fraction
among our N-rich stars. Or more plausibly, this can be explained
by orbit alteration of the present GCs by dynamical friction of the MW,
resulting in the different dynamical properties between our
NRP (GC-escaped stars) and the GCs.

We observe a significant rise up to $\sim$ 20\%, of the fraction of the N-rich stars
below [Fe/H] = --2.0, and infer that fully destroyed VMP
GCs ([Fe/H] $<$ --2.0) may have contributed in some degree to the Galactic
halo, because there are not many surviving GCs with [Fe/H] $<$ --2.0.
The overall fraction of the NRP in our giant sample is 0.4 $\pm$ 0.1\%,
which is much lower than previous studies (2 -- 3\%), which we believe is due to
the larger fraction of GSE stars and disk stars in our sample, compared to previous
studies.

The overall dynamical properties between the NRP and NNP
are dissimilar, indicating that the N-rich stars arose from diverse origins.
We conduct more detailed dynamical analysis of the N-rich
stars, dividing them into three groups according to the metallicity: the MPNR, MINR,
and MRNR groups. We find that, not only does each metallicity group exhibit
distinct dynamical properties, but the N-enhanced stars within each group show
different characteristics in their dynamical parameters.
Guided by the orbital properties of the MW GCs that are believed to be formed
in situ, we attempt to estimate the fraction of the accreted origin
in each group. We obtain an accreted fraction of 98 $\pm$ 2\%,
86 $\pm$ 4\%, and 3 $\pm$ 3\% for the MPNR, MINR, and
MRNR groups, respectively. These estimates suggest that almost all of
the MPNR stars come from fully/partially disrupted accreted metal-poor GCs,
whereas nearly all of the MRNR stars arise from fully disrupted GCs that
formed in situ in the Galactic disk system. The MINR group is
dominated by the accreted stars associated with the GSE, Sgr dwarf galaxy, and
Sequoia, but still a small fraction of them may originate from the
existing GCs. Although overall, the ex situ origins in our NRP accounts
for 64 $\pm$ 4\%, according to the selection process
of the in situ origin stars, the fraction is expected to be revised
based on future analyses of much larger samples.

To conclude, the NRP stars do not share a single common origin.
A substantial fraction of the N-rich stars appear to originate from
in situ formed GCs. However, more than 60\% of NRP stars do not
appear to be associated with stars of typical Galactic populations,
suggesting they have extragalactic origins such as GSE, Sgr dwarf galaxies,
and Sequoia, as well as from other presently unrecognized progenitors.
In this study, we attempted to
present the possible origin of distinct stellar population in each group,
based on the dynamical parameters and chemistry ([Fe/H] and [Mg/Fe]).
However, because the dynamical and chemical properties of the stars in each group are
somewhat overlapped with each other and the chemical information is limited to clearly
decompose the distinct stellar populations, more detailed
chemical-abundance analysis of the stars in each group is absolutely required, and their
abundance patterns of several elements should be compared with
those of the existing GCs and stars of other dwarf galaxies to disentangle
their diverse origin. In line with this, we plan to carry out high-resolution
spectroscopic follow-up observations for some of our N-rich stars.

\acknowledgments
We thank an anonymous referee for a careful review of this paper,
which has improved the clarity of its presentation.
Y.S.L. acknowledges support from the National Research Foundation (NRF) of
Korea grant funded by the Ministry of Science and ICT (NRF-2021R1A2C1008679).
Y.S.L. also gratefully acknowledges partial support for his visit to the University
of Notre Dame from OISE-1927130: The International Research Network for Nuclear Astrophysics (IReNA),
awarded by the US National Science Foundation. Y.K.K. acknowledges support from Basic Science
Research Program through the NRF of Korea funded by the Ministry of
Education (NRF-2021R1A6A3A01086446). T.C.B. acknowledges partial support for
this work from grant PHY 14-30152; Physics Frontier Center/JINA Center for the Evolution
of the Elements (JINA-CEE), awarded by the U.S. National Science Foundation.

Funding for the Sloan Digital Sky Survey IV has been provided by the
Alfred P. Sloan Foundation, the U.S. Department of Energy Office of
Science, and the Participating Institutions.

SDSS-IV acknowledges support and resources from
the Center for High Performance Computing  at the University of Utah. The SDSS
website is www.sdss.org.

SDSS-IV is managed by the Astrophysical Research Consortium
for the Participating Institutions of the SDSS Collaboration including
the Brazilian Participation Group, the Carnegie Institution for Science,
Carnegie Mellon University, Center for Astrophysics | Harvard \&
Smithsonian, the Chilean Participation Group, the French Participation Group,
Instituto de Astrof\'isica de Canarias, The Johns Hopkins
University, Kavli Institute for the Physics and Mathematics of the
Universe (IPMU) / University of Tokyo, the Korean Participation Group,
Lawrence Berkeley National Laboratory, Leibniz Institut f\"ur Astrophysik
Potsdam (AIP), Max-Planck-Institut f\"ur Astronomie (MPIA Heidelberg),
Max-Planck-Institut f\"ur Astrophysik (MPA Garching),
Max-Planck-Institut f\"ur Extraterrestrische Physik (MPE),
National Astronomical Observatories of China, New Mexico State University,
New York University, University of Notre Dame, Observat\'ario
Nacional / MCTI, The Ohio State University, Pennsylvania State
University, Shanghai Astronomical Observatory, United
Kingdom Participation Group, Universidad Nacional Aut\'onoma
de M\'exico, University of Arizona, University of Colorado Boulder,
University of Oxford, University of Portsmouth, University of Utah,
University of Virginia, University of Washington, University of
Wisconsin, Vanderbilt University, and Yale University.

The Guoshoujing Telescope (the Large Sky Area Multi-
Object Fiber Spectroscopic Telescope, LAMOST) is a National
Major Scientific Project which is built by the Chinese Academy
of Sciences, funded by the National Development and Reform
Commission, and operated and managed by the National
Astronomical Observatories, Chinese Academy of Sciences.

\begin{table*}[t]
\centering
\tiny
\renewcommand{\tabcolsep}{2pt}
\caption{List of IDs for N--rich Stars}
\label{tab1}
\begin{tabular}{r|r|r}
\toprule
ID~~~~~~~~~~~~~~~~~~           &ID~~~~~~~~~~~~~~~~~~           &ID~~~~~~~~~~~~~~~~~~           \\
\midrule
LAMOST~J000007.71+450217.7~  &  SDSS~J084931.26+563045.4~  &    SDSS~J151021.91+505013.8~  \\
LAMOST~J000019.59+533825.5~  &  SDSS~J095223.17+535436.3~  &    SDSS~J151443.23+060657.0~  \\
SDSS~J000150.47+284510.7~  &    SDSS~J095836.19+421757.7~  &    SDSS~J151519.74+082851.5~  \\
LAMOST~J000350.99+533248.5~  &  SDSS~J100419.85+025552.5~  &    LAMOST~J152413.93--021757.6~  \\
LAMOST~J000459.21+490514.2~  &  SDSS~J100607.70+115538.8~  &    SDSS~J152624.59+503349.0~  \\
LAMOST~J012016.08+533733.7~  &  SDSS~J100747.12+014348.6~  &    SDSS~J160327.57+173348.6~  \\
SDSS~J012108.81+395314.7~  &    SDSS~J102826.26+175448.9~  &    SDSS~J160531.97+044031.1~  \\
LAMOST~J012350.31+570326.5~  &  SDSS~J104005.94+454033.9~  &    SDSS~J160709.23+044712.7~  \\
SDSS~J014236.78--091334.8~  &   SDSS~J104730.17+221139.1~  &    SDSS~J160743.57+055029.2~  \\
SDSS~J014604.43+142639.4~  &    SDSS~J104928.34+472825.1~  &    SDSS~J161535.29+311750.0~  \\
SDSS~J022754.91+231139.6~  &    SDSS~J105403.30+482802.1~  &    LAMOST~J161746.22+242527.3~  \\
SDSS~J023303.33+230841.3~  &    SDSS~J105843.49+152902.5~  &    SDSS~J164715.46+341627.5~  \\
LAMOST~J023724.85+393802.9~  &  SDSS~J111406.40+404741.2~  &    LAMOST~J164736.87--054159.8~  \\
LAMOST~J024634.61+392504.3~  &  SDSS~J112602.86--074026.6~  &   SDSS~J165513.24+420300.2~  \\
SDSS~J024857.28+011442.0~  &    SDSS~J113026.47--004127.4~  &   SDSS~J165540.67+114924.1~  \\
SDSS~J025402.11--000718.8~  &   SDSS~J114443.02+073005.1~  &    SDSS~J165758.38+291440.4~  \\
SDSS~J025933.62+382457.3~  &    SDSS~J120101.27+125047.6~  &    SDSS~J165818.11+210330.1~  \\
SDSS~J031118.42+053248.2~  &    SDSS~J120321.15+134836.1~  &    LAMOST~J170657.57+320645.1~  \\
SDSS~J031358.15+174232.9~  &    SDSS~J120836.77+391054.9~  &    SDSS~J172556.84+081101.8~  \\
LAMOST~J033341.67+295134.2~  &  SDSS~J121024.09--011051.9~  &   SDSS~J172615.14+334012.6~  \\
LAMOST~J041612.46+505937.3~  &  SDSS~J121226.54+210123.1~  &    SDSS~J173052.92+333303.1~  \\
LAMOST~J042244.82+421623.6~  &  SDSS~J121228.73+105512.0~  &    SDSS~J173706.90+334120.3~  \\
SDSS~J044209.56+220126.6~  &    SDSS~J122428.12+045337.0~  &    SDSS~J174257.32+245017.3~  \\
LAMOST~J045052.80+481849.6~  &  SDSS~J123055.37+003428.2~  &    LAMOST~J174842.61+225103.7~  \\
SDSS~J050358.50--035445.5~  &   SDSS~J123205.79+243911.2~  &    LAMOST~J180132.38+072643.4~  \\
LAMOST~J052018.92+362128.4~  &  SDSS~J124829.56+290621.9~  &    SDSS~J181146.33+242210.7~  \\
LAMOST~J052730.80+175343.4~  &  SDSS~J125030.95+094751.7~  &    SDSS~J183725.43+411133.3~  \\
LAMOST~J053109.08+501417.5~  &  SDSS~J125249.85--013746.8~  &   SDSS~J184947.27+183643.2~  \\
LAMOST~J053119.44+183419.5~  &  SDSS~J125311.24+103948.8~  &    SDSS~J192123.01+615917.8~  \\
LAMOST~J053158.79+193844.3~  &  SDSS~J130017.60+203702.4~  &    SDSS~J200837.14--103618.6~  \\
LAMOST~J060742.47--034804.1~  & SDSS~J130224.56+294930.1~  &    SDSS~J204914.05+003145.0~  \\
LAMOST~J061423.84+325649.6~  &  SDSS~J130348.68+502648.1~  &    SDSS~J211830.64--062007.0~  \\
LAMOST~J061955.84+023338.1~  &  LAMOST~J130953.85--012145.2~  & SDSS~J211957.83--055835.8~  \\
LAMOST~J063206.58+205410.4~  &  SDSS~J131958.14--001551.9~  &   SDSS~J221334.14--072604.1~  \\
LAMOST~J064124.29+274954.1~  &  SDSS~J132109.65+412929.2~  &    SDSS~J223752.75+131427.6~  \\
LAMOST~J064348.15+250115.2~  &  SDSS~J132443.92+542751.2~  &    SDSS~J223825.62+142623.3~  \\
LAMOST~J064859.92+432420.9~  &  SDSS~J134143.89+282931.0~  &    SDSS~J224955.06+312228.8~  \\
LAMOST~J072050.58+044121.7~  &  SDSS~J134245.48+282445.3~  &    SDSS~J233305.69+004541.5~  \\
SDSS~J072157.19+381756.7~  &    SDSS~J134612.66+464651.1~  &    LAMOST~J233520.88+352011.6~  \\
SDSS~J073136.00+165109.6~  &    SDSS~J140849.44+535803.5~  &    SDSS~J233703.57+084740.2~  \\
LAMOST~J080417.90+060550.6~  &  SDSS~J145121.97+401638.9~  &    SDSS~J233906.83+073805.4~  \\
SDSS~J080735.71+071805.4~  &    SDSS~J145134.72+415036.4~  &    SDSS~J233942.28+141615.6~  \\
LAMOST~J081214.46+030750.8~  &  SDSS~J145320.16+094750.3~  &    SDSS~J234033.39+152620.4~  \\
SDSS~J081344.87+662517.4~  &    SDSS~J145500.26+235832.0~  &    SDSS~J234323.62--093251.7~  \\
SDSS~J083935.93+560913.7~  &    SDSS~J145613.98--000906.9~  &   SDSS~J234425.60--101739.1~  \\
SDSS~J084032.28+221825.9~  &    SDSS~J150721.66+454006.8~  &    SDSS~J235238.21+140745.1~  \\
\bottomrule
\end{tabular}
\end{table*} 

\end{document}